\documentclass[preprint2]{aastex}

\usepackage{natbib}
\usepackage{natbib,natbibspacing}
\usepackage{graphicx}
\usepackage{epstopdf}
\usepackage[applemac] {inputenc}
\usepackage{txfonts}
\usepackage[margin=10pt,font=small,labelfont=bf, labelformat=simple]{caption}

\slugcomment{Accepted for publication in ApJ}

\shorttitle{The Pair Fraction of Massive Galaxies at 0 $ \le z \le $ 3}
\shortauthors{Man et al.}

\begin{document}


\title{The Pair Fraction of Massive Galaxies at 0 $ \le z \le $ 3}


\author{Allison W.S. Man\altaffilmark{1} and Sune Toft \altaffilmark{1}and Andrew W. Zirm \altaffilmark{1}}
\affil{Dark Cosmology Centre, Niels Bohr Institute, University of Copenhagen, Denmark}
\email{allison@dark-cosmology.dk, sune@dark-cosmology.dk, azirm@dark-cosmology.dk}
\author{Stijn Wuyts \altaffilmark{2}}
\affil{Max-Planck-Institut f\"ur Extraterrestrische Physik, Germany}
\email{swuyts@mpe.mpg.de}
\and
\author{Arjen van der Wel\altaffilmark{3}}
\affil{Max-Planck Institut f\"ur Astronomie, Germany}
\email{vdwel@mpia.de}




\begin{abstract}
Using a mass-selected ($M_{\star} \ge 10^{11} M_{\odot}$) sample of 198 galaxies at $0 \le z \le 3.0$ with \textit{HST}/NICMOS $H_{160}$-band images from  the COSMOS survey, 
we find evidence for the evolution of the pair fraction above $z \sim 2$, an epoch in which massive galaxies are believed to undergo significant structural and mass evolution.
We observe that the pair fraction of massive galaxies is 0.15 $\pm$ 0.08 at 1.7 $\le z \le$ 3.0,
where galaxy pairs are defined as massive galaxies having a companion of flux ratio from 1:1 to 1:4 within a projected separation of 30 kpc.
This is slightly lower, but still consistent with the pair fraction measured previously in other studies, 
and the merger fraction predicted in halo-occupation modelling.
The redshift evolution of the pair fraction is described by a power law \textbf{$F(z) = (0.07 \pm 0.04) \times (1+z)^{0.6 \pm 0.5}$}.
The merger rate is consistent with no redshift evolution, however it is difficult to constrain due to the limited sample size and the high uncertainties in the merging timescale.
Based on the merger rate calculation, we estimate that a massive galaxy undergoes on average $1.1 \pm 0.5$ major merger from $z = 3$ to 0.
The observed merger fraction is sufficient to explain the number density evolution of massive galaxies, but insufficient to explain the size evolution.
This is a hint that mechanism(s) other than major merging may be required to increase the sizes of the massive, compact quiescent galaxies from $z \sim 2$ to 0.
\end{abstract}


\keywords{galaxies: evolution --- galaxies: formation --- galaxies: high-redshift}



\section{Introduction}

The sizes of quiescent massive galaxies at $z \sim 2$ are shown to be on average 3 - 6 times smaller compared to galaxies of similar mass at $z = 0$ \citep{Daddi2005, Trujillo2006b, Trujillo2006a, Toft2007, Trujillo2007, Zirm2007, Buitrago2008, Cimatti2008, Franx2008, vanDokkum2008, Toft2009, vanDokkum2010, Targett2011}.
High-resolution cosmological simulations confirm the compactness of massive galaxies at $z \sim 2$ compared to local counterparts \citep{SommerLarsen2010}.
The question is then: what are the physical processes that drive the drastic size evolution of massive quiescent galaxies between $0 < z < 2$?

At $z \sim 2$, merging is an important process for the evolution of galaxies,
in terms of mass \citep{vanDokkum2010} and size:
\citet{Khochfar2006} demonstrate with their semi-analytical model that the observed redshift-size evolution of elliptical galaxies may be a consequence of the available amount of cold gas during the major merger.
Furthermore, \citet{vanderWel2009} suggest that major merging is the most important mechanism to produce massive, quiescent galaxies through studying the distribution of the projected axial ratio of galaxies from the Sloan Digital Sky Survey.
On the other hand, there is evidence from observations \citep{Bezanson2009} and simulations \citep{Naab2009} that minor mergers are more common than major mergers and could be the dominant driver for the inferred size evolution.
Most of the luminous red elliptical galaxies at $z < 1$ are assembled through gas-poor (i.e. dry) merging \citep{Bell2004, vanDokkum2005, Bell2006b}.
The high fraction ($\sim 50\%$, \citealt{Kriek2006, Kriek2008, Williams2009}) of massive galaxies at $z \sim 2$ that are quiescent and have old stellar populations suggests that dry mergers may be common since that epoch until $z = 0$.
However, it is likely that dry mergers can only account for a factor of $\sim 2$ of growth in size from $z \sim 2$ to 0 \citep{Nipoti2009}.

Additionally, gas-rich mergers have been shown to drive gas towards the central supermassive blackholes and possibly trigger the active galactic nuclei, releasing enough energy to expel the gas and thereby quenching star formation \citep{DiMatteo2005}.
The gas inflow can also enhance star formation and even fuel starbursts \citep{Barnes1991, Mihos1994}.
The merging between two disk-like galaxies can form an elliptical, as predicted in simulations \citep{Toomre1972, Barnes1996}.
Although if the merging is highly dissipational, a larger degree of rotation and therefore flattening of the remnant may be expected \citep{Naab2006, Robertson2006, Wuyts2010}.
Recently, \citet{vanderWel2011} present evidence for the dominance of such disk-like morphologies in quiescent systems at $z \sim 2$.

Through studying the abundance of mergers of massive galaxies across redshift, we can place constraints on the current evolutionary model of these galaxies.
Substantial work exists in the literature regarding the merger fraction of galaxies at $z < 1.2$: 
merger samples can either be constructed via pair selection \citep{Zepf1989, Carlberg1994, LFevre2000, Patton2000, Lin2004, Kartaltepe2007, Lin2008, Bundy2009, Robaina2010} or morphological selection \citep{LFevre2000, Conselice2003, Lotz2008, McIntosh2008, Heiderman2009, Jogee2009}.
Kinematic evidence suggests that not all irregular morphologies at high redshift are related to mergers \citep{FSchreiber2009, FSchreiber2011}.
Hence, it is not straightforward to identify mergers through morphological classification and we focus on using pair counts as a probe for merging activity in this Paper.
It has been challenging to establish large samples of pairs of massive galaxies at $z > 1$.
High-resolution near-infrared (NIR) imaging is required to probe the rest-frame optical emission from the stellar populations.
Large-area NIR surveys have only begun recently (e.g., the CANDELS survey \citep{Grogin2011, Koekemoer2011}, the 3D-HST survey \citep{vanDokkum2011}).
There are only few spectroscopically confirmed mergers at $z > 1.5$ \citep{Shapiro2008, Law2011}.
Attempts to constrain the pair fraction at higher redshifts are limited to targetted observations \citep{Bluck2009}.
The selection criteria of massive galaxies and pairs vary across studies,
posing a challenge to make a uniform comparison of the pair fractions.

The aforementioned observations compare the observed pair fraction with the predicted merger fraction from cosmological simulations, combined with semi-analytical models (e.g., \citealt{Somerville2008, Bertone2009, Hopkins2010b}) or semi-empirical models using the halo occupation distribution \citep{Hopkins2010}.
The potential caveat is that these simulations are closely tied to observations, often normalized to reproduce the statistical observables such as the mass function, the luminosity function, and the correlation function of galaxies.

This Paper uses a sample of 198 massive ($M_{\star} \ge 10^{11} M_{\odot}$) galaxies at $0 \le z \le 3$ with high-resolution NIR imaging.
The sample is drawn from the Cosmic Evolution Survey (COSMOS) where parallel imaging from the Near-Infrared Camera and Multi-Object Spectrograph (NICMOS) onboard the \textit{Hubble Space Telescope (HST)} is available, in order to probe their rest-frame optical morphology, offering the novel opportunity to derive the pair fraction of a mass-selected sample across a wide redshift range with robust photometric redshifts (photo-$z$'s) and masses derived from spectral energy distribution (SED) fitting.

The Paper is organized as follows:
in \S\ref{sec:data} we describe the photometric catalogue, the quantities derived and the completeness of the catalogue; the selection of galaxy pairs and the correction for projection contamination are also discussed.
In \S\ref{sec:results} the results of the analysis are detailed: we compare our pair fractions with other observations and model predictions, and estimate the effect of mass completeness on the pair fraction. 
We also explore the impact of merging on the growth of the massive galaxy population: 
the merger rates are calculated and the predicted number growth is compared with the observed number densities of massive galaxies. 
Implications on our current understanding of massive galaxy formation are discussed.
The conclusions are outlined in \S\ref{sec:conclusions}.

All magnitudes are quoted in the AB system, unless otherwise stated.
A cosmology of $H_{0}$ = 70 km s$^{-1}$ Mpc$^{-1}$, $\Omega_\mathrm{M}$ = 0.3 and $\Omega_{\Lambda}$ = 0.7 is adopted throughout the Paper.

\section{Data: Catalogue and Selection} \label{sec:data}


The COSMOS field \citep{Scoville2007b} provides photometry in 30+ bands over an area of $>$ 2 deg$^{2}$, including imaging from the Advanced Camera for Surveys (ACS).
The \textit{HST}/NICMOS Camera 3 (NIC3) non-contiguously covers $\sim$ 5 \% (332 arcmin$^{2}$) of the field, with 5$\sigma$ depth of $H = 25.6$ for point sources. 
The NIC3 imaging is used with the $F160W$ filter, and the drizzled images have a pixel scale of 0.101$\arcsec$/pix and a FWHM PSF of $\sim$ 0.25$\arcsec$. 
In our analysis we use the NIC3 images from the COSMOS Archive\footnotemark reduced by James Colbert.

\footnotetext{http://irsa.ipac.caltech.edu/data/COSMOS/images/nicmos/}

\subsection{Input Catalogue}
The analysis of this Paper is based on the public COSMOS 30+ band catalogue, combined with the $H$-band photometry by \citet{Gabasch2008}, and the IRAC photometry from sCOSMOS.
The parent catalogue \citep{Ilbert2009} is selected in the $i$-band (from Subaru Suprime-Cam), where fluxes are measured within apertures of 3$\arcsec$ in diameter and has a limiting magnitude of $i < 26$.
The resulting photometric catalogue is compiled from all public data in narrow-, medium- and broad-bands covering wavelengths in UV, optical, NIR and mid-IR, and has a limiting magnitude of $K <  23.86$. 

\subsection{Derived quantities}
Photo-$z$'s are derived on all entries using the medium- and broad-band catalogue with the code EAZY \citep{Brammer2008}.
The IRAC fluxes have been downweighted by EAZY in the fitting using a template error function.
For sources with $K_\mathrm{Vega} < 22$, we model the SED in the same way as in \citet{Wuyts2007}, in order to estimate the stellar masses. 
We make use of the BC03 \citep{Bruzual2003} stellar population synthesis model with the FAST code \citep{Kriek2009}, assuming a Chabrier initial mass function (IMF), and fit the SEDs with three different star formation histories: a single stellar population without dust, an exponentially declining model with \textit{e}-folding time of 300 Myr and dust attenuation allowed to be between $A_\textrm{v}$ = 0-4, and a constant star formation model with the same range in attenuation.
We assume solar metallicity and the \citet{Calzetti2000} extinction law.

\subsection{Selection of massive galaxies in pairs}

\subsubsection{Parent sample of massive galaxies and mass completeness}\label{massive_sample}

We select galaxy pairs by searching for companions to the massive galaxies in the NICMOS parallels.
The parent sample consists of 5,299 massive galaxies of $M_{\star} \ge 10^{11}~M_{\odot}$ at $0 \le z \le 3$ in the COSMOS field.
The photo-$z$'s are required to have \verb"odds" $\ge$ 0.95 such that they have $\ge$ 95\% integrated probability of lying within $\Delta z$ = 0.2 of the estimate.
The $\chi^{2}$-value of the SED modelling is required to be less than 10.
The \verb"odds" and the $\chi^{2}$ criteria reject approximately 55\% and 17\% of all the sources at $0 \le z \le 3.0$ in the whole COSMOS catalogue, 
ensuring robustness in the photo-$z$'s and masses.

\begin{figure}[h!]
	\centering
	\includegraphics[angle=0,width=0.5\textwidth]{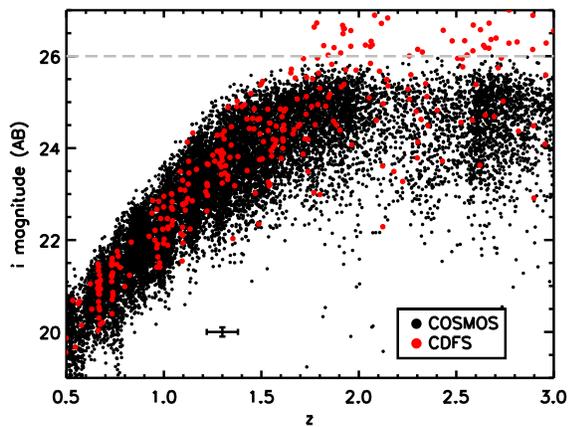}
	\caption{Mass completeness of the catalogue.
	Galaxies above our mass limit of $M_{\star} \ge 10^{11} M_{\odot}$ in COSMOS (black) and CDFS (red) are plotted.
	The gray dashed line shows the approximate depth of the COSMOS data.
	The typical uncertainties in magnitudes and photo-$z$'s are overplotted at the bottom left corner.}
	\label{fig:mass_comp}
\end{figure}

To estimate the completeness of our adopted mass limit of $M_{\star} \ge 10^{11}~M_{\odot}$ from our $i$-band selected catalogue, 
we compare the selected galaxies to the $K$-band selected FIREWORKS catalogue \citep{Wuyts2008} for GOODS-Chandra Deep Field South (CDFS), which has a deeper limiting magnitude of $i = 27$ (3$\sigma$).
As the completeness is a strong function of redshift, 
we compare the magnitude distribution of massive galaxies in COSMOS and CDFS against redshift in Figure~\ref{fig:mass_comp}.
Assuming that CDFS is 100\% complete in selecting massive galaxies, the completeness limit of COSMOS are 100\%, 75\% and 44\% for the redshift bins 0-1.7, 1.7-2.3 and 2.3-3.0 respectively.
Only 7\% and 14 \% of the massive galaxies are rejected by the \verb"odds" and the $\chi^{2}$ criteria,
so the incompleteness is mostly due to the faintness of the massive galaxies in the $i$-band.

There are 305 massive galaxies in the parent sample which have NICMOS $H_{160}$ parallels, 
but 109 of those are in the edge region of low signal to noise, 
so there are 196 galaxies with usable NICMOS imaging.

\subsubsection{Selection of galaxy pairs}\label{massive_pairs}

We run \verb"SExtractor" Version 2.8.6 \citep{Bertin1996} on the $10\arcsec \times10\arcsec$ NICMOS cutouts (or $35\arcsec \times 35 \arcsec$ for galaxies at $0 \le z \le1.0$), with parameters optimized to ensure that sources are deblended properly.
The isophotal fluxes are used to compute the $H$-magnitude of each source. 
Due to the large photometric aperture used in the COSMOS catalogue, 22 cutouts have more than one source within the 3\arcsec-aperture, where the source-confused companions have no separate entry in the catalogue.
In these cases the photo-$z$ of the companion is assigned to be the same as the primary massive galaxy, and the integrated best-fit masses are adjusted using the NICMOS $H$-band flux ratio from \verb"SExtractor".
The final massive galaxy sample consists of 198 massive ($M_{\star} > 10^{11}M_{\odot}$) galaxies at $0 \le z \le 3$, where all of them are brighter than the depth of the NICMOS imaging described in \S\ref{sec:data}.
Note that there are two more massive galaxies compared to the 196 massive galaxies mentioned in \S\ref{massive_sample}.
This is because there was source confusion in the photometry of two of the selected massive galaxies, 
and after mass correction there are two massive galaxies on each cutout (four massive galaxies in total).

\begin{figure}[h!]
	\centering
	\includegraphics[angle=0,width=0.5\textwidth]{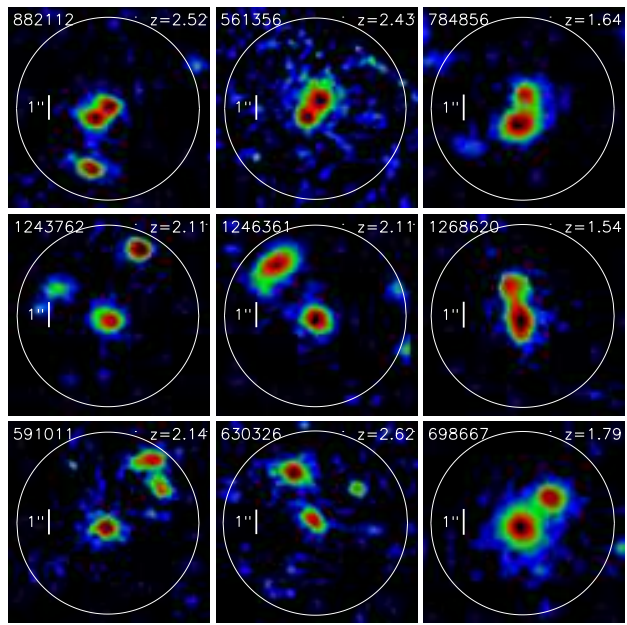}
	\caption{The NICMOS $H_{160}$ postage stamps of nine examples of the selected galaxy pairs.
	The top row shows pairs that were source-confused in the original COSMOS catalogue, but are now resolved in our analysis with the NICMOS imaging;
	the bottom two rows contain pairs that have individual entries in the catalogue.
	The IDs and photo-$z$'s of the massive galaxies are labelled on the top left and right hand corners of each panel.
	For illustrative purpose, the colour coding is scaled logarithmically and the images are smoothed by convolving with a Gaussian PSF of FWHM = 2 pixels ($0.202\arcsec$).
	The angular scale is shown with the 1$\arcsec$ vertical bar.
	The white circle overlaid on each map indicates the 30-kpc search radius around each massive galaxy  at the centre.
	}
	\label{fig:pairs_images}
\end{figure}

Galaxy pairs are selected from the massive galaxies sample using the following criteria:
(1) the massive galaxy has one or more companion within a projected separation of 30 kpc; and
(2) the $H_{160}$-flux ratio of the pair is between 1:4 to 1:1.
Imposing these criteria we find 40 massive galaxies in pairs (N$_\mathrm{obs}$), in the redshift range of $0 \le z \le 3$.
Almost all (99\%) of the massive galaxies are bright enough such that if they have 1:4 companions, the companions are brighter than the depth of NICMOS ($H = 25.6$, see \S\ref{sec:data}).
Only two massive galaxies (1\%) are fainter than $H = 25.6$, but they are retained in the sample because they have companions detected with NICMOS.
The number of galaxies having companions of flux ratio above 1:2 / 1:3 / 1:4 is 20 / 32 / 40 respectively.
Examples of the cutouts are shown in Figure~\ref{fig:pairs_images}.
We note that there are two galaxy pairs in which both of the merging galaxies are massive ($M_{\star} > 10^{11}M_{\odot}$).

\subsubsection{Correcting for chance projection} \label{projection}

Before comparing the pair fraction with model predictions and investigating its redshift evolution, it is necessary to subtract the contamination from projected galaxy pairs at different redshifts. 
We estimate the effect of chance projection by performing a Monte-Carlo simulation, assuming that there is no clustering in the sources or the massive galaxies.
All COSMOS sources are redistributed randomly over the effective (unmasked) area of the COSMOS field.
The 198 massive galaxies in the sample are also assigned random positions.
Using their photo-$z$'s we count, within an annulus of 5-kpc to 30-kpc, the number of close companions that have magnitudes down to 1:4 fainter.
We repeat the redistribution and counting for 500 realizations.
The average of the counts are taken as the expected number of galaxies in projected pairs ($< \mathrm{N_{projected}} >$) for each redshift bin, and are listed in Table~\ref{table:pf}.
It can be seen that approximately half of the observed pairs are random projections.
We also note that the correction is more significant at higher redshift.
This is because high-$z$ galaxies are fainter and the surface number counts are higher for faint galaxies, resulting in a higher probability of chance projection.
\citet{Law2011} find a similar correction for chance projection ($\sim$ 50\%) using spectroscopic redshifts available for 2874 star-forming galaxies at $1.5 < z < 3.5$.

In each redshift bin, we observe M massive galaxies and $\mathrm{N_{obs}}$ of them are in pairs.
The fraction of galaxies in pairs ($f_{p}$), or pair fraction for short, is calculated as:
\begin{displaymath}
		$$ \[ f_{p} = \frac { \mathrm{N_{obs}} -  < \mathrm{N_{projected}}> } {\mathrm{M}} \] $$
\end{displaymath}
The errors in $f_{p}$ are estimated by the Poisson uncertainties of N$_\mathrm{obs}$.

Alternatively, as photo-$z$'s are available for 68\% of the companions and all of the massive galaxies,
we can use the photo-$z$'s to reject projected pairs and identify the pairs that are physically associated.
In practice, the pairs are identified by the separation and flux ratio criteria, and additionally a photo-$z$ critierion:
if the companion has a separate COSMOS entry with reliable photo-$z$ (\verb"odds" $\ge$ 0.95), the 3$\sigma$ confidence intervals of the photo-$z$'s must overlap.
The number of massive galaxies in pairs is given by $\mathrm{N_{p}^{\prime}}$, and the pair fraction is given by $f_{p} = \mathrm{N^{\prime}_{p}}~ / ~{\mathrm{M}}$.

\section{Results and Discussions} \label{sec:results}




\subsection{Comparison with previous observations} \label{pair_z}

\begin{figure}[h!]
	\centering
	\includegraphics[angle=0,width=0.5\textwidth]{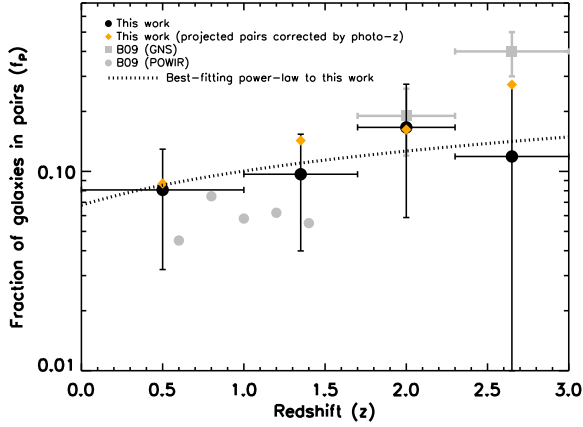}
	\caption{The redshift evolution of the pair fractions, compared to other observations. 
	The black circles denote the $f_{p}$ of our analysis, after statistically correcting for projection contamination.
	The black dotted line show the best-fitting power law to our $f_{p}$, which is of the form $F(z) = (0.07 \pm 0.04) \times (1+z)^{0.6 \pm 0.5}$.
	The orange diamonds denote our $f_{p}$, which we use an alternative approach to correct for projection contamination with the available photo-$z$'s.
	The gray squares and circles represent the $f_{p}$ of \citet{Bluck2009} using the GNS and POWIR data.
	 The horizontal bars indicate the width of each bin.
	}
	\label{fig:pf_lit}
\end{figure}

In order to examine the redshift evolution of the pair fraction, we correct for chance projections in the observed pairs in COSMOS to get $f_{p}$ in different redshift bins through our default approach, as listed in Table~\ref{table:pf} and plotted on Figure~\ref{fig:pf_lit}.
The pair fraction derived using the photo-$z$ criterion is remarkably consistent with our default approach, except at the highest redshift bin in which photo-$z$'s have higher uncertainties and therefore less constraining, but still marginally consistent.
The Kolmogrov-Smirnov test confirms that the fraction of massive galaxies in pairs is inconsistent with no redshift evolution.
We fit the observed $f_{p}$ with a power law of the form $F(z) = F(0) (1+z)^{m}$, and find the best fit parameters to be $F(0) = 0.07 \pm 0.04$ and $m = 0.6 \pm 0.5$.

\citet{Robaina2010} (hereafter R10) use the amplitude of the projected two-point correlation function of massive galaxies to estimate the pair fraction of galaxies in the COSMOS and COMBO-17 surveys at $z = 0-1.2$, 
requiring galaxy pairs to be separated by less than 30 kpc in three-dimensional space, and each galaxy to be more massive than $5 \times 10^{10}$~$M_{\odot}$.
Applying their mass limit to both galaxies in our pairs, we find a $f_{p}$ consistent with their results at $z \le 1.2$,
though we note that it is rare to find two galaxies that are both massive in a close pair.
Our sample of $\sim 200$ massive galaxies yield relatively large uncertainties in the pair fraction due to small number counts, compared to R10's sample of $\sim$18,000 massive galaxies.
The agreement ensures that our results are compatible with previously published $f_{p}$ below $z = 1.2$ \citep{Xu2004, Bell2006, McIntosh2008, Bundy2009} that are consistent with R10's $f_{p}$.

Our corrected pair fractions agree to that of \citet{Law2011} within the uncertainties, though we note that their sample are based on star-forming galaxies above $10^{10} M_{\odot}$.

\citet{Bluck2009} (hereafter B09) present a study of 82 massive galaxies with NICMOS imaging at $1.7 < z < 3$ from the GOODS NICMOS survey (GNS),
and define pairs as any galaxy within 30 kpc and within a difference of $\pm$1.5 in $H_{160}$-magnitude compared to the host massive galaxy.
They find the pair fraction to be 0.19 $\pm$ 0.07 at $1.7 < z < 2.3$ and 0.40 $\pm$ 0.10 at $2.3 < z < 3$.
The comparison is shown in Figure~\ref{fig:pf_lit}.
We also compare to the pair fraction from the POWIR survey at $z \sim 1$ from B09.
Note that the COSMOS catalogue is $i$-band selected,
whereas the GNS targets are selected using three different criteria (Distant Red Galaxies, Infrared Extremely Red Objects and BzK galaxies); \citep{Conselice2011}.
Despite the fact that the difference in the selection could potentially bias the results, our agreement with the \citet{Bluck2009} results is a strong confirmation of high pair fraction at $z \sim 2$.

\subsection{How robust is our pair fraction?} \label{robustness}

In our analysis, we quote the pair fraction based on the relative fraction of galaxies to a certain depth.
When the pair fraction is translated to a merger fraction, it is necessary to account for any systematic bias of our massive galaxy sample, i.e. whether the galaxies that we miss due to the limited $i$-band depth have the same pair fraction.
We perform a test to estimate the effect of the incompleteness on our observed pair fraction.
In \S\ref{massive_sample}, we demonstrate that the adopted mass limit is 75\% (44\%) complete in selecting massive galaxies at $1.7 \le z \le 2.3$ ($2.3 \le z \le3.0$).
If our sample of 37 massive galaxies is 75\% complete at $1.7 \le z \le 2.3$, we can estimate that in total there are $\sim$49 massive galaxies, and we miss $\sim$12 of them because of their faintness in the $i$-band.
If we assume the extreme scenarios, in which all the missed galaxies are (not) in pairs, the pair fraction in $1.7 \le z \le 2.3$ would then be 0.37 $\pm$ 0.11 (0.12 $\pm$ 0.08).
For $2.3 \le z \le 3.0$, a similar calculation yields $f_{p}$ of 0.05 $\pm$ 0.07 and 0.61 $\pm$ 0.13 at the limits.
It is apparent the conservative lower limits are within the errors of our observed $f_{p}$.

The massive galaxies fainter than $i = 26$ are likely to be at the high redshift end, and the faintness can be explained by dusty star formation or evolved stellar populations.
Using the deeper, $K$-band selected CDFS catalogue, we find that $f_{p}$ = 0.21$^{+0.26}_{-0.17}$ at $1.5 \le z \le 3.0$ for $26 < i < 27$, in the fainter regime where the COSMOS catalogue is incomplete.
This is consistent with our expectation that the missed galaxies would have a similar $f_{p}$ as observed in COSMOS for the $i$-band brighter galaxies. 
We note that the CDFS covers a smaller area than COSMOS, and hence statistical errors in the resulting $f_{p}$ are more severe. 

\subsection{The growth of the massive galaxy population through merging} \label{implications}

\subsubsection{Merger rate}\label{merger_rate}
We calculate the merger rate as $\Re(z) = f_{p} (z) n(z) \tau^{-1}$, 
where the merging timescale ($\tau$) is assumed to be $0.4 \pm 0.2$ Gyr \citep{Lotz2008b},
and the observed co-moving number density of massive galaxies is the number of massive galaxies (M) divided by the co-moving volume in that redshift range subtended by the usable area of 474 NICMOS pointings,
i.e. $n(z)$ = M$(z)$ / $V_\mathrm{co-moving}$.
The completeness limits derived in \S\ref{massive_sample} are used to correct M.
The merger rates are listed in Table~\ref{table:pf}.
As an estimate, the uncertainties of $f_{p}$, $n(z)$ and $\tau$ are approximately $69\%$, $20\%$ and $50\%$ respectively.
This yields an uncertainty of $\sim 88\%$ in the derived merger rate.
Therefore, the major merger rate, unlike the pair fraction, is consistent with no redshift evolution within the large range of uncertainties.
The characteristic time between mergers ($\Gamma$) experienced by a galaxy at a given redshift is given by $\Gamma = \tau / f_{p}$,
and we find the best fit to its redshift evolution to be $\Gamma = 12 (1 + z)^{-1.6}$.
By integrating $\Gamma$ over our redshift range (see Equation (6) of B09),
we estimate that a galaxy experiences $N_{m} = 1.1 \pm 0.5$ major mergers on average from $z=3.0$ to $z=0$,
consistent with B09's $N_{m} = 1.7 \pm 0.5$ within the large uncertainties.

\subsubsection{Number density evolution}
\begin{figure}[h!]
	\centering
	\includegraphics[angle=0,width=0.5\textwidth]{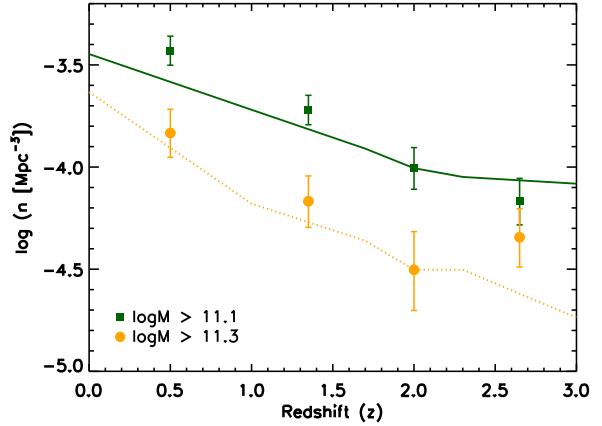}
	\caption{The redshift evolution of number density of massive galaxies.
	The filled symbols are the observed co-moving number density of massive galaxies from our sample, with mass limits shown in the legend.
	The lines represent the predicted number growth using the observed number density of close pairs, after correcting for projected pairs using photo-$z$.
	The lines are normalized to the observed number density at $z = 2$.}
	\label{fig:numdens}
\end{figure}

The mass function (MF) of galaxies is altered by mergers.
If the merger fraction at different epochs is known, one can translate it into the evolution of the massive galaxy population assuming a merging timescale ($\tau = 0.4 \pm 0.2$ Gyr in this Paper).
We estimate the number of newly created massive galaxies using the selected galaxy pairs:
for each pair, we calculate the remnant mass as the sum of the SED masses of the galaxies in the pair.
In the rare case (4 pairs) where the SED mass is not available for the companion galaxy because there is no corresponding entry in the catalogue, 
we use the flux ratio and the SED mass of the primary massive galaxy to estimate the remnant mass.
Here we have assumed that the $H_{160}$ flux ratio corresponds to the mass ratio, and we verify the assumption by finding consistent remnant masses using the flux and the mass ratios for the remaining pairs.
The number of newly created massive galaxy (N$_\mathrm{created}$) in each redshift bin is calculated by counting the galaxies that cross the mass limit after merging.
The merger-induced increment in the co-moving number density ($\Delta$, in units of Mpc$^{-3}$) is given by:
\begin{displaymath}
		$$ \[ \Delta = \frac{\mathrm{N_{created} } \times t_\mathrm{elapsed}}{ V_\mathrm{co-moving} \times  \tau} \] $$
\end{displaymath}
where $t_\mathrm{elapsed}$ is the time elapsed within the redshift bin.
Our selected galaxy pairs consists of primary galaxies of $M_{\star} \ge 10^{11} M_{\odot}$, with companions of flux ratio down to 1:4.
Therefore, the remnant mass would be at least $1.25 \times 10^{11} M_{\odot}$ $(logM = 11.1)$.
In the case of an equal-mass merger, the remnant mass will be $2\times 10^{11} M_{\odot}$ $(logM = 11.3)$.
Normalizing the number density of massive galaxies to the observation at $z=2$,
the results are compared with the observed number density of massive galaxies above these mass limits, 
as shown in Figure~\ref{fig:numdens}.
Considering the $\sim$ 0.2 dex uncertainty in the number density growth due to counting statistics,
the slope of the number growth is remarkably consistent with the observed number density.
As the highest redshift bin ($z > 2.3$) is only 44\% complete, the projected number growth is highly uncertain. 
The agreement between our estimated merger-induced number density growth and the observed number density supports the idea that major mergers are sufficient to explain the number density evolution of massive galaxies from $z \sim 2.3$ to 0.

One potential caveat of this test is that mergers of galaxies less massive than $10^{11} M_{\odot}$ are not included, due to mass incompleteness of the catalogue.
Another caveat is the assumption that no new stars are formed in the merging, which is only valid for dry merging.
This can result in an underestimation of the number density growth of $log(M) > 11.1$ galaxies, where equal-mass mergers of two galaxies of down to $log(M) > 10.8$ could contribute to the number density.
The number density evolution of massive galaxies depends on several factors:
a merger between less massive galaxies can create a massive galaxy above the mass limit;
on the other hand, if two massive galaxies merge, the number of massive galaxies would be reduced;
the merging timescale is closely related to the growth rate of the massive galaxies.
The buildup of the massive galaxies can be better constrained with larger samples of galaxies down to lower masses and higher redshifts,
which will be feasible with the upcoming surveys. 
The study of number density evolution is complimentary to the mass density evolution \citep{Conselice2007} and the mass evolution of a fixed number density sample across redshift \citep{vanDokkum2010}, 
in tracing the buildup of massive galaxies.
Our observed number densities show an agreement to \citet{vanDokkum2010}'s finding that the stellar mass of massive galaxies double since z = 2.
A precise measurement of the contribution of mergers to the number density evolution requires accurate determination of the merging timescale, and is beyond the scope of this Paper.

If massive galaxies undergo $\sim 1.1 \pm 0.5$ major merger between $0 < z < 3$ and this is sufficient to explain the number density evolution, 
then this hints that major merging can be ruled out as the main mechanism for puffing up the sizes of massive, compact and quiescent galaxies from $z \sim 2$ to 0, as this size evolution requires 2-3 major mergers \citep{Bezanson2009, Toft2009}.

\subsubsection{Comparison with models} \label{compare_model}
\begin{figure}[h!]
	\centering
	\includegraphics[angle=0,width=0.5\textwidth]{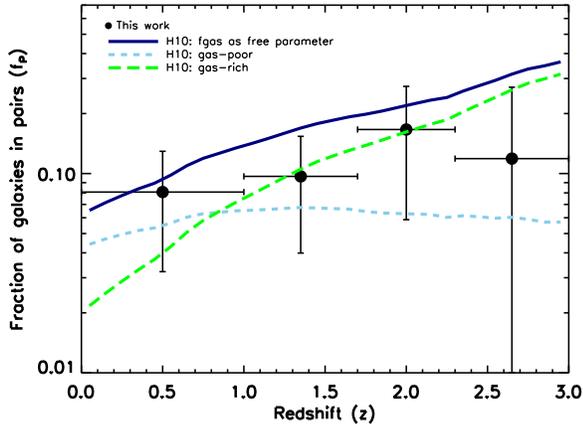}
	\caption{A plot similar to Figure~\ref{fig:pf_lit} that compares our pair fractions to model predictions.
	 The colour lines are the predicted $f_{p}$ for pair-selected samples from the merger rate calculator of \citet{Hopkins2010} assuming different sets of gas fraction.}
	\label{fig:pf_model}
\end{figure}

To understand how our observations fit into the current understanding of galaxy formation in a cosmological context, 
we compare our observed pair fraction to the expected merger fraction in pair-selected samples computed from the ``merger rate calculator" (MRC) developed by \citet{Hopkins2010} (hereafter H10).
H10 use a halo occupation model to track merger history,
according to the merger trees constructed from the Millennium Simulation \citep{Fakhouri2008}.
The galaxy-galaxy merger rate is determined by convolving the distribution of galaxies in halos with the dynamical timescale.
Assuming a merging timescale of $0.35 \pm 0.15$ Gyr \citep{Lotz2008b}, the merger rate is then converted to a merger fraction.
Using a simplified fitting function, the MRC predicts the merger fraction as a function of galaxy mass, gas fraction, redshift and mass ratio.
We compute the galaxy-galaxy merger fraction at $0 \le z \le 3.0$ for galaxies of stellar mass between $10^{11}$ and $10^{12}$ $M_{\odot}$, the range of masses of our massive galaxies sample, and of mass ratio down to 1:4.
The average gas fraction of the pair, defined as $f_\mathrm{gas}$ =  $M_\mathrm{gas}$ / ( $M_\mathrm{gas}$ + $M_\mathrm{\star}$ ), is a free parameter in the model.
Direct measurements of the gas mass fraction of massive star-forming galaxies at z $\sim$ 1 and $z \sim 2$ give 34\% and 44\% respectively \citep{Tacconi2010}; and 50\%-65\% for similar systems at $z \sim 1.5$ in another study \citep{Daddi2010}.
These systems are considered evidence for very gas-rich systems at those epochs,
hence we select the critical gas fraction ($f_\mathrm{gas}^{\star}$) to be 20\% to differentiate gas-poor ( 0 $\le$ $f_\mathrm{gas}$ $\le$ $f_\mathrm{gas}^{\star}$ ) and gas-rich ($f_\mathrm{gas}^{\star}$ $\le$ $f_\mathrm{gas}$ $\le$ 1) mergers.
Then we multiply the merger fraction, as a function of redshift, by two to get the predicted fraction of galaxies in pairs to compare with our observations in Figure~\ref{fig:pf_model}.
To investigate the importance of dry merging (i.e. nearly dissipationless mergers), we overplot the the gas-poor and gas-rich pair fraction for comparison.

Considering that the systematic uncertainties in the predicted pair fraction are larger than a factor of two,
our observed pair fraction is consistent with the prediction of H10's model.
The number of gas-rich mergers are sufficient to explain the number of observed pairs at $1 \le z \le 3$.
Gas-poor mergers are predicted to be more frequent than gas-rich mergers below $z = 1$ (see H10), and are required to explain the observed pair fraction.
H10's model predicts 2.1 major mergers per galaxy from $z=3$ to $z=0$, which is almost twice of our result and is apparent from the predicted pair fraction in Figure~\ref{fig:pf_model}.
Our result is also low compared to $\sim$ 1 major merger per galaxy at $0 < z <1.5$ for $M_{\star} \ge 10^{10.8} M_{\odot}$ (mass limit converted from Salpeter into Chabrier IMF for comparison) predicted by \citet{Drory2008}, who estimate the contribution of merging by subtracting the effect of mass-dependent star formation from the galaxy stellar MF.
The discrepancy is mostly due to the difference in mass limit, confirmed by a similar value of 1.1 major merger predicted by \citet{Hopkins2010}'s MRC if we use \citet{Drory2008}'s mass limit.
To reproduce their results of $\sim 2$ major mergers we need to use a merging timescale of 0.25 Gyr, which is lower, yet still within the uncertainties of \citet{Lotz2008b}'s range of merging timescale for pairs having projected separation up to 30 kpc.
The expected number of major majors from \citet{Drory2008} and \citet{Hopkins2010} are based on observations of the MF,
whereas our result is a direct measurement of the pair fraction converted into the number of major mergers using the merging timescale and the number density of massive galaxies.
This illustrates the need to better constrain the merging timescale, and to improve the understanding of how merging alters the MF, in order to push merger rate measurements to higher accuracy.

\section{Conclusions} \label{sec:conclusions}
We have quantified the pair fraction of 198 massive ($M_{\star} \ge 10^{11} M_{\odot}$) galaxies at $0 \le z \le 3$ from COSMOS with NICMOS parallels. 
Our findings provide a confirmation of the evolution of pair fraction from $z = 3$ to $z = 0$, in agreement with previous observations \citep{Bluck2009} and predictions from halo-occupation modelling \citep{Hopkins2010}.
Gas-rich mergers are sufficient to explain the observed pair fraction from $z = 3$ to 1; below $z = 1$ gas-poor mergers are also needed.
The fraction of massive galaxies observed to be in pairs is 0.15 $\pm$ 0.08 from 1.7 $\le z \le$ 3.0.
The redshift evolution of the pair fraction is described by a power law $F(z) = (0.07 \pm 0.04) \times (1+z)^{0.6 \pm 0.5}$.
The merger rate is consistent with no redshift evolution, though the uncertainties in pair counts and merging timescale restrict the ability to conclusively constrain the merger rate.
On average, a massive galaxy undergoes $\sim 1.1 \pm 0.5$ major merger from $z = 3$ to 0, assuming a merging timescale of 0.4 Gyr.
Using the inferred merger fraction, we are able to reproduce the observed number density of massive galaxies since $z \sim 2.3$.
This implies that major merging can account for the number density evolution of the massive galaxies, but other mechanisms such as minor merging may be required to explain the size evolution of the massive, compact quiescent galaxies at $z \sim 2$.



\acknowledgments{}
The Dark Cosmology Centre is funded by the Danish National Research Foundation. S. Toft and A. W. Zirm gratefully acknowledge support from the Lundbeck Foundation.
The authors thank the referee, Thomas Targett, for useful comments that helped improve the paper.
A. W. S. Man thanks Kinwah Wu, Anna Gallazzi, Steen Hansen and Thomas Greve for insightful discussions.

{\it Facility: HST (NICMOS)} 

\clearpage
\begin{deluxetable}{ccccc} 
	\tablecolumns{4}
	\tablewidth{0pc}
	\tablecaption{Pair fraction and merger rate across redshifts}
	\tablehead{   
	  \colhead{\textbf{Redshift range}} &
 	  \colhead{\textbf{No. of massive galaxies (in pairs) }} &
	  \colhead{\textbf{Expected no. of galaxies }} &
 	  \colhead{\textbf{Pair fraction}} &
  	  \colhead{\textbf{Merger rate $\Re(z)$} } \\
	  \colhead{} &
	  \colhead{ \textbf{M (N$_\mathrm{obs}$)}} &
	  \colhead{ \textbf{in projected pairs $<$N$_\mathrm{projected}>$}} &
	  \colhead{} &
	( $\times 10^{4}$ Gpc$^{-3}$ Gyr$^{-1}$ )
	}
	\startdata
	$0 \le z \le 1.0$ & 69 (8) & 2.4 & 0.08 $\pm$ 0.05  & 12.0 \\ 
	$1.0 \le z \le 1.7$ & 70 (12) & 5.2 & 0.10 $\pm$ 0.06 & 7.7 \\ 
	$1.7 \le z \le 2.3$ & 37 (12) & 5.9  & 0.17 $\pm$ 0.11 & 9.2 \\  
	$2.3 \le z \le 3.0$ & 22 (8) & 5.4 & 0.12 $\pm$ 0.15 & 5.6 \\
	\textbf{1.7 $\le$ \textbf{\textit{z}} $\le$ 3.0} & \textbf{59 (20)} & \textbf{11.3} & \textbf{0.15 $\pm$ 0.08} \\ 
	\enddata
	\label{table:pf}
\end{deluxetable}

\clearpage

\clearpage
\bibliographystyle{apj} 

\end{document}